\documentstyle[epsfig,aps,pra]{revtex}
\begin{document}

\newcommand{\ket}[1]{| #1 \rangle}
\newcommand{\bra}[1]{\langle #1 |}
\newcommand{\braket}[2]{\langle #1 | #2 \rangle}
\newcommand{\proj}[1]{| #1\rangle\!\langle #1 |}

\twocolumn[\hsize\textwidth\columnwidth\hsize\csname
@twocolumnfalse\endcsname

]

\centerline {\Large \bf Simple Pulses for Universal}
\centerline {\Large \bf Quantum Computation with a}
\centerline {\Large \bf Heisenberg ABAB Chain} 
\bigskip

\centerline {{\bf Simon C. Benjamin} (s.benjamin@qubit.org)}
\smallskip
\centerline {Centre for Quantum Computation, Clarendon}
\centerline {Laboratory, University of Oxford, OX1 3PU, UK.}
\bigskip
{\bf Recently Levy \cite{levy} has shown that quantum computation can be performed using an ABAB.. chain of spin-1/2 systems with nearest-neighbor Heisenberg interactions. Levy notes that all necessary elementary computational `gates' can be achieved by using spin-resonance techniques involving modulating the spin-spin interaction strength at high frequency. Here we note that, as an alternative to that approach, it is possible to perform the elementary gates with simple, non-oscillatory pulses.} 

\bigskip 

Consider a pair of independent (pseudo-)spin $1\over 2$ systems, with transition energies $A$ and $B$. Now suppose that these systems may be coupled by a Heisenberg-type interaction, so that the Hamiltonian is:

 ${\hat H}=-{A\over 2}{\hat \sigma_A^z}\otimes{\hat I}_B-{B\over 2}{\hat I}_A\otimes{\hat \sigma_B^z}+J{\hat {\underline \sigma}}_A\otimes{\hat {\underline \sigma}}_B$. 

\noindent Here 
$\hbar=1$, subscripts $A$ and $B$ refer to the $2\otimes2$ subspace of the corresponding system, \{${\hat \sigma}^x$, ${\hat \sigma}^y$, ${\hat \sigma}^z$\} are the Pauli matrices, and ${\underline {\hat \sigma}}\equiv{\underline i}{\hat \sigma}^x+{\underline j}{\hat \sigma}^y+{\underline k}{\hat \sigma}^z$. We will consider the dynamics of the system as the magnitude of $J$ is abruptly switched between steady values (the square wave-case). The other simple pulse shapes, such as the more realistic Gaussian form, will have comparable effects.

The dynamics of the constant-$J$ system are easy to establish by diagonalising ${\hat H}$. It is convenient to add a (physically meaningless) global energy-shift term $J{\hat I}_A\otimes{\hat I}_B$ - this provides a slight simplification to the matrix form of $H$:

${\hat H}\rightarrow\left( 
\begin{array}{cccc}
-\Omega+K & 0 & 0 & 0 \\ 
0 & \omega & K & 0 \\ 
0 & K &  -\omega & 0 \\ 
0 & 0 & 0 & \Omega+K
\end{array}
\right) $ 

\noindent in the basis $\{\ket{00} ,\ket{01} ,\ket{10},\ket{11}$\}.
Here $K\equiv2J$, $\Omega\equiv{1\over 2}(A+B)$ and $\omega\equiv{1\over 2}(A-B)$.
Since ${\hat H}$ is already diagonal in the \{$\ket{00}$,$\ket{11}$\} subspace, we can concentrate on the \{$\ket{01}$,$\ket{10}$\} subspace. Following Levy, we will identify this as the subspace of a single logical qubit, writing $\ket{01}\equiv\ket{0}_L$ and $\ket{10}\equiv\ket{1}_L$. Diagonalisation is straightforward:

${\hat H}_L\rightarrow\left( 
\begin{array}{cc}
\omega & K \\ 
K &  -\omega 
\end{array}
\right) = 
{\bf R}^\dagger \left( 
\begin{array}{cc}
\omega^\prime & 0 \\ 
0 &  -\omega^\prime 
\end{array}
\right) {\bf R}=\omega^\prime {\bf R}^\dagger {\hat \sigma}_z {\bf R}$

\noindent in basis \{$\ket{0}_L$,$\ket{1}_L$\}, where $\omega^\prime=(\omega^2+K^2)^{1\over 2}$ and

${\bf R}=\left( 
\begin{array}{cc}
{\rm cos}\ {\theta \over 2} & {\rm sin}\ {\theta \over 2} \\ 
-{\rm sin}\ {\theta \over 2} & {\rm cos}\ {\theta \over 2} 
\end{array}
\right)$\ \ with $\theta={\rm ArcTan}({2J\over \omega})$.

\noindent The effect on the logical qubit of applying $J=J_0$ for a period $t$ is therefore given (in the basis \{$\ket{0}_L$,$\ket{1}_L$\}) by 

$
{\hat U}(t)={\rm exp}(-i{\hat H}_Lt)={\rm exp}\Big(-i\omega^\prime t{\bf R}^\dagger \sigma_z{\bf R}\Big)$

$\ \ \ \ \ \ ={\bf R}^\dagger \left( 
\begin{array}{cc}
{\rm e}^{-i\omega^\prime t} & 0\\ 
0 &{\rm e}^{i\omega^\prime t}
\end{array}
\right)
{\bf R} $

\noindent 
In order to understand this is terms of the Bloch sphere, we employ the operator
${\hat R}_{\underline n}(\psi)\equiv{\rm cos}({\psi\over 2}){\hat I}-i\ {\rm sin}({\psi\over 2})(n_x{\hat \sigma}_x+n_y{\hat \sigma}_y+n_z{\hat \sigma}_z)$. This represents a rotation on the sphere by $\psi$ radians about the axis specified by unit vector ${\underline n}={\underline i}n_x+{\underline j}n_x+{\underline ik}n_x$ \cite{NC}. Then we find that our ${\hat U}(t)={\hat R}_{\underline \theta}(2\omega^\prime t)$ with ${\underline \theta}\equiv {\rm cos}\ \theta\ {\underline i}+{\rm sin}\ \theta\ {\underline j}$, i.e. the effect of applying $J=J_0$ for time $t$ is a rotation by $2 \omega^\prime t$ about an axis in the z-x plane. (All rotations here are in the lab frame; the rotating frame is considered later).

\begin{figure}
\centerline{\epsfig{file=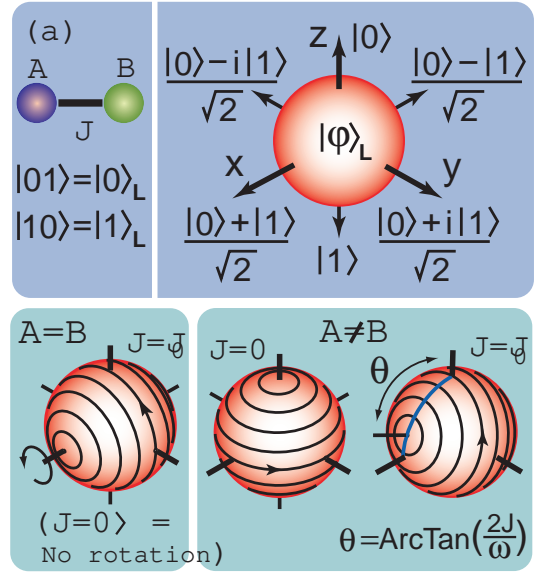,width=7.cm}}
\vspace{0.2cm}
\caption{(a) The two spin-1/2 systems and the corresponding Bloch sphere for the logical qubit. (b) Schematics showing the flow of states over time, depending on $\omega\equiv {1\over 2}(A-B)$ and $J$.}
\label{figure1}
\end{figure}

Let us assume that $\omega\equiv {1\over 2}(A-B)$ is fixed (as in the case, for example, that our physical qubits $A$ and $B$ are electron spins in quantum dots of different local g-factor in a {\em static} global $B$-field). We will also assume that $J$ cannot be switched from positive right though to negative \cite{noJinvert}. Then we can never-the-less vary the axis of rotation by choosing the magnitude of $J$. If $J=0$ then the rotation is ${\hat R}_{\underline k}(2\omega t)$, i.e. a simple rotation about the z-axis. With $J>0$ we have a rotation about an axis lying in the z-x plane at an angle $\theta={\rm ArcTan}(2J/\omega)$ to the z-direction. To achieve a rotation about an axis close to the x-direction, we would therefore require a very large $J$ value (infinite for a pure x-rotation). This is impractical, but we can instead synthesize a pure y-axis rotation by a sequence of more modest rotations. For example, since
${\hat R}_{\underline j}(2\theta)={\hat R}_{\underline k}(\pi){\hat R}_{\underline \theta}(\pi)$, we can generate any ${\hat R}_{\underline j}(0\leq\psi\leq 2\pi/3)$ provided that the range of available $J$ is $0\leq J\leq{\sqrt 3}\ \omega$. Moreover we can concatenate such pairs of rotations in order to achieve any ${\hat R}_{\underline j}(0\leq\psi\leq 2\pi)$ - a maximum of 3 pairs will suffice. Fig 2 shows two such pairs being concatenated to produce ${\hat R}_{\underline j}(\pi)$.

\begin{figure}
\centerline{\epsfig{file=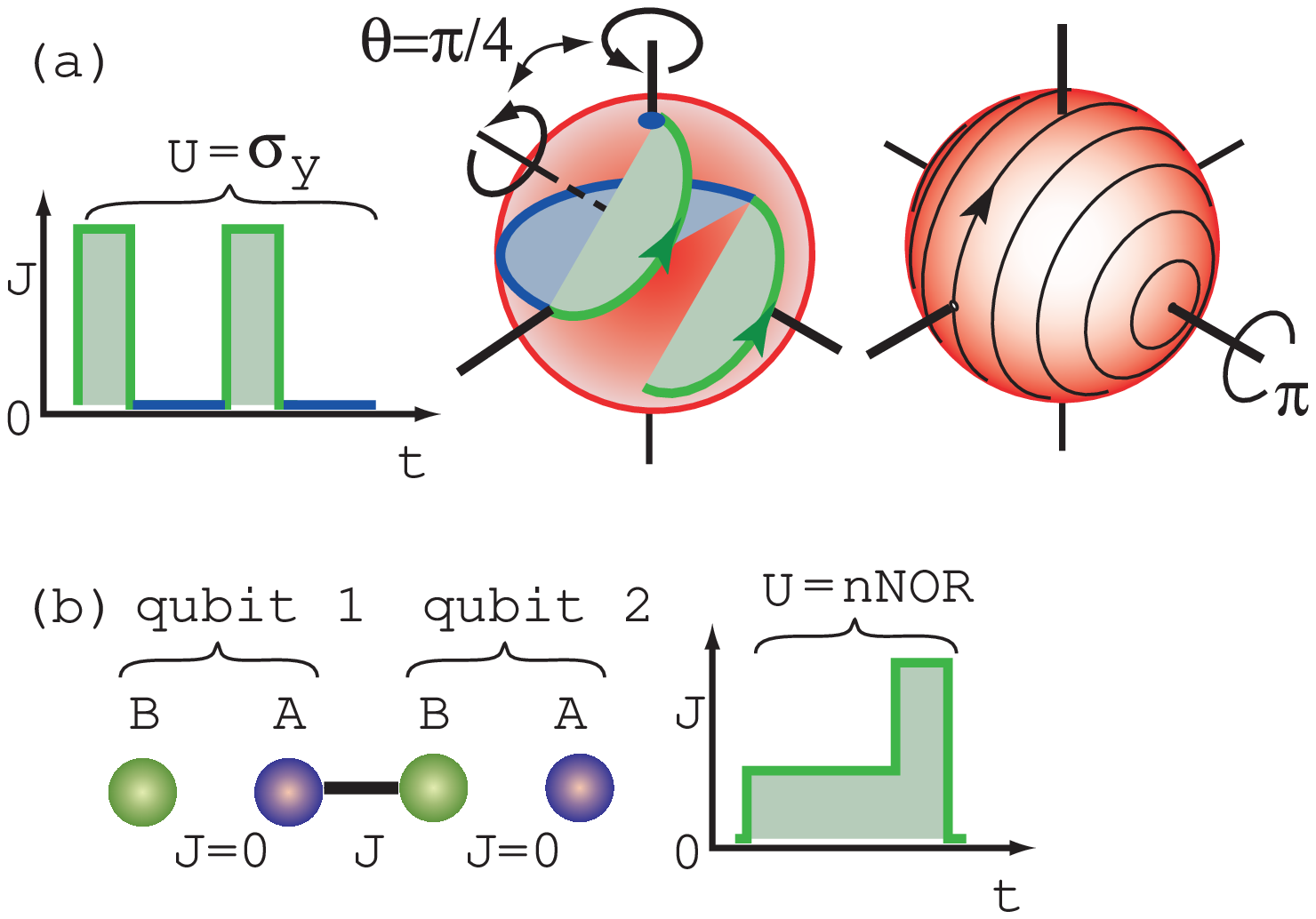,width=7.8cm}}
\vspace{0.2cm}
\caption{(a) A sequence of four steps to synthesize ${\hat R}_{\underline j}(\pi)$.} (b) Two steps suffice for a certain two-qubit gate.
\label{figure1}
\end{figure}

Given that we can achieve pure z-rotations and pure y-rotations, we can use the sequence ${\hat R}_{\underline k}(\alpha) {\hat R}_{\underline j}(\beta) {\hat R}_{\underline k}(\gamma)$ to synthesize (up to a meaningless global phase)
the general single-qubit transform 

${\hat G}=\left( 
\begin{array}{cc}
e^{-i(\alpha/2+\gamma/2)}{\rm cos}\ {\beta \over 2} & -e^{i(-\alpha/2+\gamma/2)}{\rm sin}\ {\beta \over 2} \\ 
e^{i(\alpha/2-\gamma/2)}{\rm sin}\ {\beta \over 2} & e^{i(\alpha/2+\gamma/2)}{\rm cos}\ {\beta \over 2} 
\end{array}
\right)$. 

\noindent This formal construction therefore corresponds to a maximum of 7 steps for any single qubit gate ($1+6+1=8$, but we may amalgamate the last two, since both are z-rotations). In practice, there will be shorter sequences for any given operation. For example, the important Hadamard transform corresponds to just a single step (e.g. applying $J=2\omega_0$ for time $t=\pi/(2\sqrt 2 \omega_0)$). The time requirement for the ${\hat R}_{\underline j}(\pi)$ rotation shown in Fig. 2(a) is probably quite typical - it is $\pi (1+{\sqrt 2})/(2\omega)$. 

One might object that since the {\em other} qubits in the computer are also (presumably) represented by an AB pair, these qubits will have performed a z-axis rotation ${\hat R}_{\underline k}(2\omega \tau)$ whilst we were performing ${\hat G}$ on our target qubit. We should take these rotations into account, i.e. we should really be working in the rotating frame of a passive qubit. A naive method (not the most efficient) for achieving this is to supplement our ${\hat G}$ sequence with a rotation $R_\theta(2\pi)$, which has no net effect in the lab frame but takes time $\tau^\prime=\pi/\omega^\prime$. With an appropriate choice of $\theta$  $(\Rightarrow \omega^\prime)$ the total gate time $\tau$ is then such that $\omega \tau = 2n \pi$, so that the `other' qubits have experienced zero net rotation. More efficiently, one would incorporate this consideration into the process of deriving the optimal short rotation sequence for ${\hat G}$.

The above analysis therefore demonstrates that any single qubit-gate can be efficiently performed on the logical qubit via by a short sequence of fixed $J$ values. It is straightforward to extend this approach to produce a particular two-qubit gate  which, together with our universal single-qubit gate, will form a  complete set of gates for computation. Consider an BABA section of a quantum computer, and suppose that two logical qubits are represented in this section, one in the first BA pair and one in the second (see Fig 2(b)). Now suppose that the interaction is ``off'' between all spins except the middle AB pair (which spans the two logical qubits). With an appropriate short sequence \cite{seq} of non-zero J values, we can produce the net effect

${\hat U}_{gate}=i\left( 
\begin{array}{cccc}
1 & 0 & 0 & 0 \\
0 & 1 & 0 & 0 \\
0 & 0 & -1 & 0 \\
0 & 0 & 0 & 1 \\
\end{array}
\right).{\hat U}_0$

\noindent in the basis $\{\ket{00} ,\ket{01} ,\ket{10},\ket{11}$\} of the central two spins. Here ${\hat U}_0$ denotes the time evolution that {\em would have occurred} if the interaction had simply been off for the whole period. Thus the effect  (up to a meaningless global phase of $i$) is to introduce a phase of $-1$ conditional on central spin-pair AB being in state $\ket{10}$. Remembering that the logical qubits on the two BA pairs are represented as $\ket{01}\equiv\ket{0}_L$ and $\ket{10}\equiv\ket{1}_L$, this condition translates to both logical qubits being in state $\ket{0}_L$. Our transformation is therefore a two-qubit gate comparable to the ``nAND'' gate, except that it singles-out $\ket{0}_L\ket{0}_L$ rather than $\ket{1}_L\ket{1}_L$. We might therefore describe this gate as a ``nNOR''.

As a final remark, it is worth noting that although the above approach does not require the $\omega\equiv(A-B)/2$ parameter to be varied, never-the-less such an ability would be advantageous. In particular, it would be useful if $\omega$ could be switched to zero, because this would then allow the SWAP operation to be performed with a single pulse, and on a time scale limited only by the maximum strength of $J$. Any one-dimensional computer based on nearest-neighbor interactions must spend much its time simply moving qubits around, therefore efficient performance of SWAP is very
desirable. One might imagine a quantum dot implementation where the B-field has a cycle involving being `off' for a period of the time (during which qubits are moved around), before being pulsed to a large value in order to allow general one and two-qubit gates as described above.

To conclude, we have explicitly shown that one can perform universal computation in the system described by Levy using only simple fixed values of $J$. This scheme, with its relatively modest set of physical requirements, is a strong candidate architecture for solid state quantum computing.

The author wishes to thank Ernesto Galvao and Jeremy Levy for useful conversations.


\begin{references}
\bibitem{seq} One such sequence for the given matrix consists of just two steps as follows: for time $t=2\pi a_+/\omega_0$ apply $J=2\omega_0(1/a_-^2+1)^{1\over 2}$, then for time $t=2\pi a_-/\omega_0$ apply $J=2\omega_0(1/a_-^2-1)^{1\over 2}$. Here $a_{+/-}=(5 +/- {\sqrt 7})/8$.
\end{references}
\end{document}